\documentclass[12pt]{article}

\usepackage{a41}
\usepackage{epsfig}
\newcounter{my}

\newcommand{\la}[1]{\label{#1}}
\newcommand{\re}[1]{\ (\ref{#1})}
\newcommand{\nn}{\nonumber}
\newcommand{\ed}{\end{document}}
\newcommand{\be}{\begin{equation}}
\newcommand{\ee}{\end{equation}}
\newcommand{\ba}{\begin{eqnarray}}
\newcommand{\ea}{\end{eqnarray}}
\newcommand{\baz}{\begin{eqnarray*}}
\newcommand{\eaz}{\end{eqnarray*}}
\newcommand{\bb}{\begin{thebibliography}}
\newcommand{\eb}{\end{thebibliography}}
\newcommand{\ct}[1]{${\cite{#1}}$}

\newcommand{\bi}[1]{\bibitem{#1}}

\begin{document}

\sloppy


\mbox{}
\begin{center}
{\Large \bf  A New Mechanism for Single Spin
Asymmetries in Strong Interactions}\\
\vspace{5mm}
N.I. Kochelev \\
\vspace{5mm}
{\small\it
 Bogoliubov Laboratory of Theoretical Physics,\\
Joint Institute for Nuclear Research,\\
 Dubna, Moscow region, 141980 Russia}\\
\end{center}
\begin{abstract}
It is shown that the contribution of the instantons to the
fragmentation of quarks leads to   the
appearance of a imaginary part in   diagrams
of  quark-quark scattering at large transfer momentum.
The imaginary part comes from the analytical continuation
of the instanton amplitudes from Euclidean to Minkowsky
space-time and reflects quasiclassical origin of instanton solution of
QCD equations of motion.
This phenomenon and
instanton--induced quark spin-flip  give a new nonperturbative mechanism
for the  observed  anomalous single-spin asymmetries
in hadron-hadron and lepton-hadron  interactions.
\end{abstract}

\vskip 1.0cm
\leftline{Pacs: 12.39-x, 13.60.Hb, 14.65-q, 14.70Dj}
\leftline{Keywords:  spin, pion, quark, gluon, instanton, asymmetry.}
\vspace{0.5cm}

\section{Introduction}
\vspace{1mm}
\noindent

The  explanation of  the large observed single-spin
asymmetries (SSA)
at large energies and
momentum  transfers  in many inclusive and exclusive processes
is one of most longstanding and outstanding
problems in QCD \ct{E704},\ct{Kri}.
From naive point of view one can expect that with increasing of the energy
and the momentum transfer the role of the quark spin in strong
interaction should become smaller. At the same time the experimental data
 on spin-dependent cross sections reveal the opposite tendency:
the spin asymmetries do not disappear  at large energy
showing in fact an anomalous growth with increasing of the momentum transfer.

Apart from our understanding of the origin of the large spin
effects in QCD , the investigation of the fundamental mechanism
which is responsible  for SSA
is also very important in view of future spin measurements at
Brookhaven (RHIC-Spin Collaboration), CERN (COMPASS) and
 DESY (HERA-$\vec{N}$).

Within leading order of perturbative QCD it is impossible to obtain the
large SSA because their values  should be proportional to the current quark
masses  and decrease with energy and momentum transfer \ct{pqcd}.
Additional suppression factors are those related with the loop integration
generating the imaginary part of the amplitude and an extra power of
$\alpha_s$.

The several attempts to explain the observed SSA have been undertaken.
In \ct{efremov} it was mentioned that twist-3 contribution can be important
to explain this puzzle.
Recently in \ct{qiusterman} the convolution formula for single-spin
asymmetry which include twist-3 quark-gluon correlation function
has been obtained and  the single-spin
asymmetries  for the pion production have been estimated.
The main problem of this approach is  unknown spin-dependent
twist-3 distribution functions which in general are the functions of
two variables and present  the nonperturbative   part of
the convolution formula together with rather well known  twist-2
distribution functions of the partons in nonpolarized nucleon.
These twist-3 distribution functions should be either extracted from other
experiments or calculated within some nonperturbative approach.

There are several phenomenological approaches which also take into
account the nonperturbative effects on single-spin asymmetries \ct{models}.
Some of them are using  assumptions about quark transverse  momenta
in the distribution function (Sivers effect \ct{sivers}) or in the
quark fragmentation function (Collins effect \ct{collins}).
In \ct{anselmino}  the attempt to combine these two
mechanisms  for SSA has been made.

However all of these approaches are based on the phenomenological
ways to introduce nonperturbative effects into SSA problem.
The most of the parameters of these models were obtained from the fit
of the available SSA experimental data, therefore the predictable power
of such models is rather low.

In this Letter a new mechanism for single-spin asymmetries in strong
interaction is suggested.

This mechanism is based on the existence in the QCD vacuum of the strong
nonperturbative fluctuations of gluon fields, so-called instantons
(see recent review \ct{shuryak}).
The  instanton model of QCD vacuum describes very well not only the
main properties of the vacuum state e.g. the values of the different
quark and gluon condensates but it is also rather successful in the description
of the hadron spectroscopy (see recent review \ct{shuryak} and
references therein). Recent results of the lattice QCD \ct{lattice}
confirm the importance of the instantons in QCD vacuum.

The importance of the instantons for spin physics is related to their
specific role in chiral structure of QCD vacuum.
Thus the  instantons describe subbarrier transitions between various
classical minima of the QCD potential which correspond to different values of
the axial vector charge. The  changing of the value of the axial
vector charge due to instanton transition leads simultaneously to the
quark chirality flip.  In \ct{koch1} was mentioned that the quark chirality
 flip induced by instantons  may give the natural explanation of the anomalous
spin effects in strong interaction. In particular the instanton solution
of famous ``spin crisis'' \ct{rev} has been suggested \ct{koch2}.

We will show below that the
instanton  leads to a precise behavior of effective quark-instanton
vertices as functions of the incoming quark virtualities, which will
be responsible for the magnitude of the SSA. More specifically
an imaginary part arises for  time-like virtualities of the  quark
in the diagrams induced by instantons, which
is not suppressed at high energy and whose contribution enhances
significantly the SSA.
\section{Single-spin asymmetries in $\pi$ meson production and
instantons}
\vspace{1mm}
\noindent

Let us estimate the instanton contribution to the SSA for hadron
production
in quark-quark scattering.  For definiteness we study
$\pi^+$ meson production in the scattering of
two u-quarks, one of them  transversely polarized.
Our goal is to explain the large SSA in the fragmentation
region of the polarized quark at high energies. In this
kinematical  regime only   the diagrams of Fig.1 can contribute
significantly \footnote{We assume that instanton induced
quark-quark interaction determines the  strength of $\pi qq$ vertex.
This is one of the important consequences of instanton model
(see \ct{shuryak} and references therein).}.
\begin{figure}[htb]
\centering
\epsfig{file=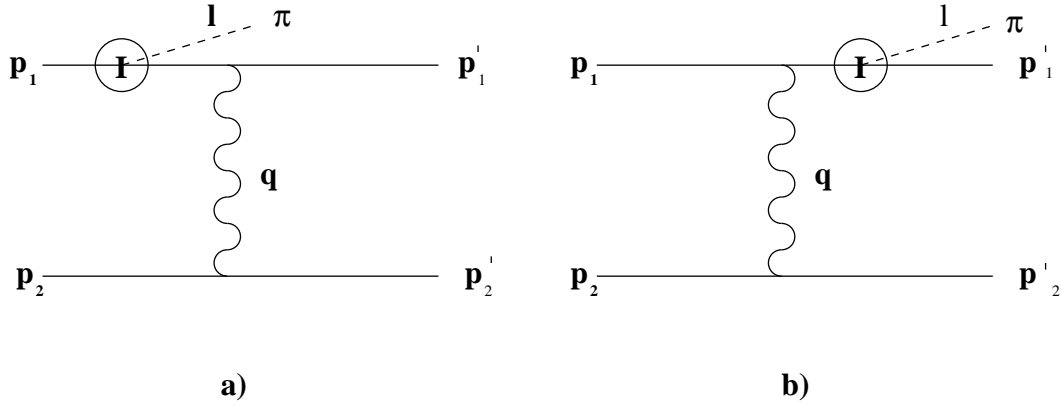,width=14cm}
\vskip 1cm
\caption{\it The  contribution of the instanton to the
amplitude for $\pi^+$ production in the fragmentation region of the
polarized quark in the scattering of two u-quarks.
 The label $I$ denotes
instanton.}
\end{figure}
The method for calculating these diagrams
is standard (see for example \ct{kuraev}).
The single spin asymmetry  can be written in the
following  form \ct{am}
\be
A=\frac{2Im(\Phi_5^*(\Phi_1+\Phi_3))}{|\Phi_1|^2+|\Phi_3|^2+4|\Phi_5|^2},
\la{as}
\ee
where we have neglected the contribution which comes from double spin-flip
amplitudes $\Phi_2$ and $\Phi_4$, which  are suppressed by factor
$(m_q/\sqrt{S})$ with respect to the leading contributions; $m_q$ is
the quark mass and $S=(p_1+p_2)^2 $. The
helicity amplitudes entering \re{as} are
\begin{equation}
\Phi_1=M_{+,+;+,+},{\  } {\ } \Phi_3=M_{+,-;+,-}, \Phi_5=M_{++;-+}.
\label{aml}
\end{equation}
By using for the  gluon polarization tensor in Fig.1 its high-energy limit
\ct{kuraev}
\be
D^{\mu\nu}=g_\bot^{\mu\nu}+\frac{2}{S}(p_1^\mu p_2^\nu+p_1^\nu p_2^\mu)
\approx\frac{2}{S}(p_1^\mu p_2^\nu+p_1^\nu p_2^\mu),
\la{prop5}
\ee
the matrix element of the reaction
\be
u(p_1)\uparrow+u(p_2)\rightarrow d(p_1^\prime)+u(p_2^\prime)+\pi^+(l)
\la{reac}
\ee
is given by
\be
\tilde{M}=\frac{2g_s^2g_{\pi^+qq}}{Sq^2}\bar u(p_2^\prime)\hat{p}_1
t^au(p_2)\bar u(p_1^\prime)p_2^\mu O_\mu u(p_1),
\la{ma1}
\ee
where
$g_s$ is the strong coupling constant, $g_{\pi^+qq}$ is the $\pi^+$--quark
coupling constant due to the instanton,
and
\be
Q_\mu=\gamma_5(\frac{\hat{l}\gamma_\mu}{d_1}F(k_1^2)-
\frac{\gamma_\mu \hat{l}}{d_2}F(k_2^2)).
\la{oper}
\ee
In order to obtain \re{oper} the equations of motion has been used.
In \re{oper} $F(k^2)$ is form factor related with the finite size
of instanton, $d_{1,2}$ are quark propagators in Fig.1, namely,
\be
d_1=(p_1^\prime+l)^2-m_q^2, {\ } {\ } d_2=(p_1-l)^2-m_q^2.
\la{prop1}
\ee
In principle \re{ma1} should include an integral of the
density of instantons $n(\rho) $ over the instanton size $\rho$.
However for estimating the SSA here we use the simple version
of the instanton liquid model, $n(\rho)=n_0\delta(\rho-\rho_c)$,
with fixed instanton size,  $\rho_c=1.6 GeV^{-1}$.
This model gives a good description of
the hadronic properties and is very suitable for obtaining
estimates\ct{shuryak}. In  \re{as} the density
of the instantons is in the numerator and the denominator. Therefore
in the ratio it cancels.
The structure   in color space of all the helicity amplitudes
in \re{as} is the same. Therefore we can omit
all global factors of \re{ma1} in the ratio as well.
Thus it is enough to consider the matrix element
\be
M=\frac{1}{Sq^2}\bar u(p_2^\prime)\hat{p}_1u(p_2)
\bar u(p_1^\prime)p_2^\mu Q_\mu u(p_1).
\la{ma2}
\ee
In the high energy limit its very suitable to use Sudakov variables
\ba
l&=&x_F\tilde{p}_1+\beta_l\tilde{p}_2+l_\bot\nn\\
p_1^\prime&=&(1-x_F)\tilde{p}_1+\beta_1\tilde{p}_2+{p_{1\bot}}^\prime\nn\\
q&=&\alpha\tilde{p_2}+\beta_q\tilde{p}_1+q_\bot,
\la{sud}
\ea
where
\be
\tilde{p}_1=p_1^\mu-\frac{m_q^2}{S}p_2^\mu, {\ } {\ }
\tilde{p}_2=p_2^\mu-\frac{m_q^2}{S}p_1^\mu,
\la{p12}
\ee
and $\tilde{p}_1^2=\tilde{p}_2^2\rightarrow 0$ at $S\gg m_q^2$.
In this limit,  in the bottom part of
the diagrams in Fig.1, due to \re{ma2}  we have conservation of
 quark helicity
\ba
\left.\{\bar u(p_2^\prime)\hat{p}_1u(p_2)\right.\}
_{\lambda\lambda^\prime}\approx
\delta_{\lambda_2\lambda_2^\prime}S
\la{bottom}
\ea
and therefore for helicity amplitude  we have
\ba
M_{\{\lambda_1,\lambda_2;\lambda_1^\prime,\lambda_2^\prime\}}=
-\delta_{\lambda_2\lambda_2^\prime}
\frac{1}{q^2\beta_q} \left.\{\bar u(p_1^\prime)q_\bot^\mu
O_\mu u(p_1)\right.\}_{\lambda_1,\lambda_1^\prime}
\la{ma3}
\ea
where the current conservation condition for the quark current in the
top line in Fig.1, $q.J^{1}=0$,  has been used and
\be
\beta_q=\frac{m_q^2x_F^2+l_\bot^2+x_F(q_\bot^2-2l_\bot.q_\bot)}{Sx_F(1-x_F)}.
\nn
\ee
By using the on-shell conditions for outcoming quarks,
$p_1^\prime=p_2^\prime=m_q^2$,  and neglecting
the mass of pion, $l^2=0$, one can easily obtain the following
expressions
for the quark propagators, $d_{1,2}$, and  quark virtualities
in the intermediate states,  $k_{1,2}$, of Fig.1,
\ba
d_1&=&\frac{m_q^2x_F^2+(x_Fq_\bot-l_\bot)^2}{x_F(1-x_F)} {\ } {\ }
d_2=-\frac{m_q^2x_F^2+l_\bot^2}{x_F}\la{prop}\\
k_1^2&=&\frac{m_q^2x_F+(x_Fq_\bot-l_\bot)^2}{x_F(1-x_F)} {\ } {\ }
k_2^2=-\frac{l_\bot^2-x_F(1-x_F)m_q^2}{x_F}.
\la{vir}
\ea
By using of the identity
\be
\gamma_\nu\gamma_\mu=g_{\mu\nu}+i\sigma_{\mu\nu}
\nn
\ee
one can write
\be
\bar u(p_1^\prime)\gamma_5\hat{l}\hat{q}_\bot=
-(q_\bot.l_\bot)\bar u(p_1^\prime)\gamma_5 u(p_1)+
l_\nu q^\mu_\bot\bar u(p_1^\prime)i\gamma_5\sigma_{\mu\nu}u(p_1).
\la{fo}
\ee
The matrix elements of operators in \re{fo} for the different helicity
states at high energy are known:
\ba
[\bar u\gamma_5 u]_{\lambda_1\lambda_1^\prime }
&\approx& \delta_{\lambda_1\lambda_1^\prime}2m_q+
\delta_{\lambda_1,-\lambda_1^\prime}|q_\bot-l_\bot|,\\\nn
[\bar ui\gamma_5l_\nu q^\mu_\bot u]_{\lambda_1\lambda_1^\prime}&\approx&
 -\delta_{\lambda_1,-\lambda_1^\prime}\frac{l_\bot^2|q_\bot|}{2x_F}
\la{tt}
\ea
Therefore the final result for helicity amplitudes in \re{as}
is
\ba
\Phi_{\lambda_1,\lambda_1^\prime;\lambda_2,\lambda_2^\prime}&=&
\frac{\delta_{\lambda_2,\lambda_2^\prime}}{q^2\beta_q}\left.\{
{\delta_{\lambda_1,-\lambda_1^\prime}}
[(q_\bot.l_\bot)|q_\bot-l_\bot|(\frac{F(k_1^2)}{d_1}-\frac{F(k_2^2)}
{d_2})\right.
\nn\\
&+&\left.\frac{l_\bot^2|q_\bot|}{2x_F}(\frac{F(k_1^2)}{d_1}+\frac{F(k_2^2)}{d_2})]
+\delta_{\lambda_1,\lambda_1^\prime}2m_q(q_\bot.l_\bot)(
\frac{F(k_1^2)}{d_1}-\frac{F(k_2^2)}{d_2})\right.\}.
\la{final}
\ea
The main feature of the
instanton--induced form factor $F(k^2)$ in \re{final} is its
nontrivial dependence on
the virtualities of the incoming quarks into the instanton vertex.
In the general case of the on-shell pion and off-shell quarks
with virtualities $k_1^2$ and $k_2^2 $ the effective quark-pion vertex has
the following form
\footnote{ The origin of  the effective quark-pion vertex
within instanton model is the famous  t'Hooft's four-quark interaction
related to the quark zero-modes in instanton field \ct{thooft}.}
\be
g_{\pi qq}(k_1^2,k_2^2)=g_{\pi qq}F(k_1^2)F(k^2),
\ee
where $F(k_i^2)$ is related to the Fourier transformed zero-mode of the quark
in the instanton field in a singular gauge (see \ct{shuryak})
\be
F(k^2)=-x\frac{d}{dx}\left.\{I_0(x)K_0(x)-I_1(x)K_1(x)\right.\},
\la{form}
\ee
where $x=\rho\sqrt{k^2}/2$.

The instanton
is a classical solution of the QCD equations of motion in
{\it Euclidean space-time} which is characterized by its size $\rho$ in
{\it this} space-time.  Therefore to obtain  the result for the cross section
the analytical continuation of instanton amplitudes to the {\it physical}
Minkowsky space-time should be done. This continuation should be
performed in the careful way because the instanton-induced amplitudes
have the cut at the quark virtuality $k^2=0$ (see below). It is this
cut which is responsible for the appearance of the imaginary part which is
needed for SSA.

To eliminate the imaginary part it is more suitable
to use a  good approximation for this form factor
which gives the correct behavior for quark zero-mode at
$k^2\rightarrow\infty$
\be
F(k^2)\approx\frac{1}{1+\rho^3(\sqrt{k^2})^3/6},
\la{form1}
\ee
 For a space--like value of the quark virtuality in the intermediate
state in Fig.1a we have $k_2^2<0$ and therefore in {\it Minkowsky}
space--time the form factor becomes
\be
F(k_2^2)\approx\frac{1}{1+\rho^3(\sqrt{|k_2|^2})^3/6}.
\la{form2}
\ee
At the same time for time--like virtuality in Fig.1b $k_1^2>0$ in
{\it Minkowsky} space--time one obtains an {\it imaginary} part in
form factor.
\be
F(k_1^2)\approx\frac{1}{1+i\rho^3(\sqrt{|k_1|^2})^3/6}.
\la{form3}
\ee
It is very well known that to get significant single spin asymmetry
one must have both quark-spin flip  and a large imaginary part in the
amplitude
\footnote{ It is interesting that only nonspin-flip amplitude in
\re{final} is proportional to the quark mass. In pQCD approach \ct{pqcd}
we have opposite situation e.g. spin-flip amplitude is proportional
to the quark mass. The difference comes from the additional quark
spin-flip at quark-pion vertex in Fig.1.}.
Eqs.\re{final} and \re{form3} show that in the instanton induced
diagrams have both of these components.
The  SSA  is proportional to the interference of diagrams in Fig.1
One can interpret the contribution from first diagram (Fig.1a).
as a Sivers effect \ct{sivers} in the quark distribution function and
the contribution from the second diagram (Fig.1b) as a contribution to
quark fragmentation function, the so--called Collins effect \ct{collins}.

To obtain the final result for the asymmetry  one should integrate
in \re{as} the numerator and the denominator over $q_\bot$ and regularize in
some way the gluon propagator at small $q^2$. The usual procedure
(see for example \ct{kuraev}) is to substitute in the  gluon
propagator,
$q^2\rightarrow-(q_\bot^2+\mu^2)$, where we have used
for the infrared regulator $\mu\approx\Lambda_{QCD}\approx m_q\approx 0.35$ GeV.
The value of $m_q=0.35$ GeV is  the contituent quark mass within
instanton liquid model \ct{shuryak}.

The result of the calculation of the  SSA
is presented in Fig.2 as a function of $x_F$ and in Fig.3
as a function of both $x_F$ and $p_\bot$.
\begin{figure}[htb]
\centering
\begin{minipage}[c]{7cm}
\centering
\epsfig{file=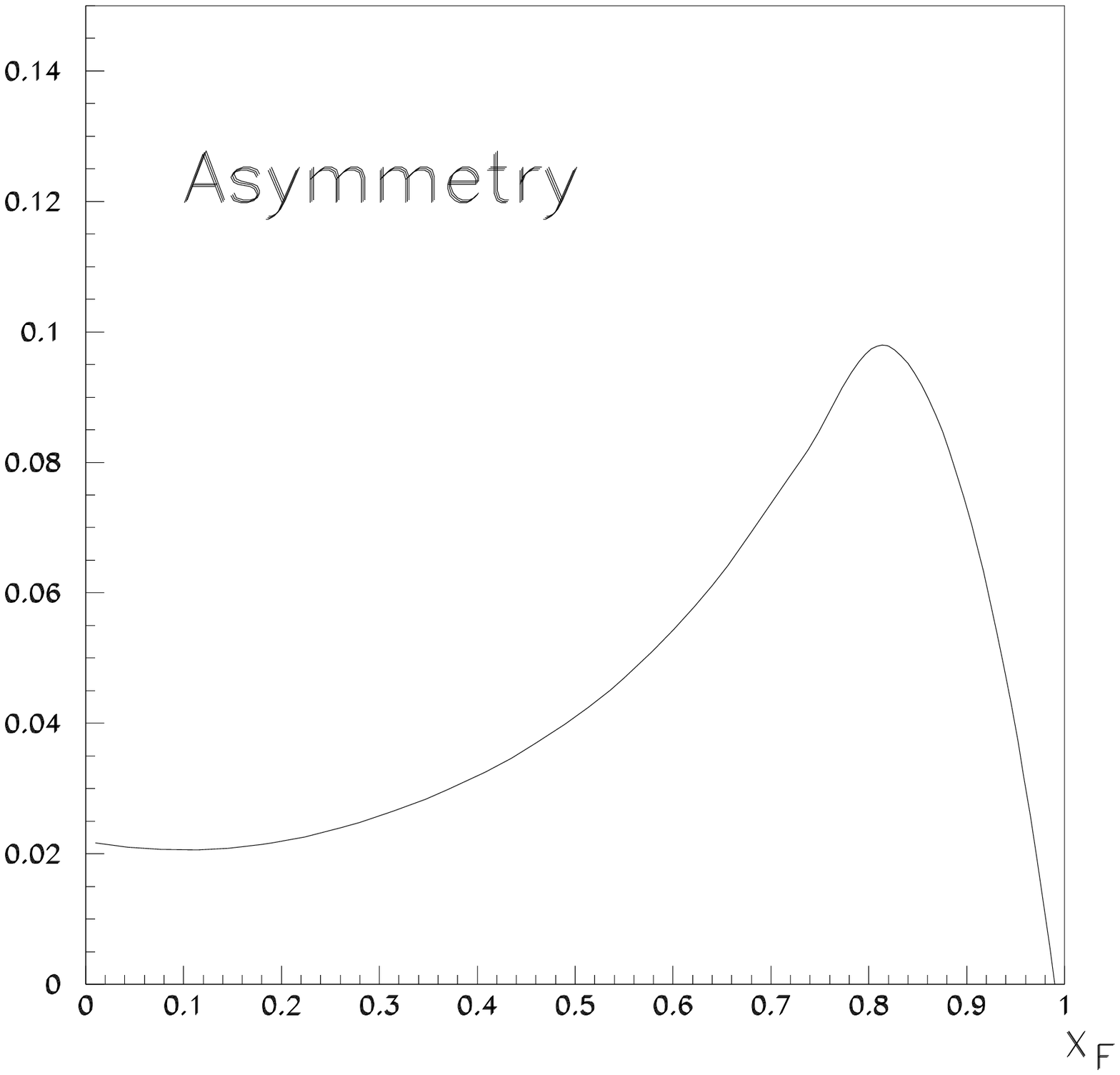,width=6cm}
\caption{\it The instanton contribution to  the  single
spin asymmetry for pion production as a function of $x_F$.}
\vspace*{1cm}
\end{minipage}
\hspace*{1cm}
\begin{minipage}[c]{7cm}
\centering
\epsfig{file=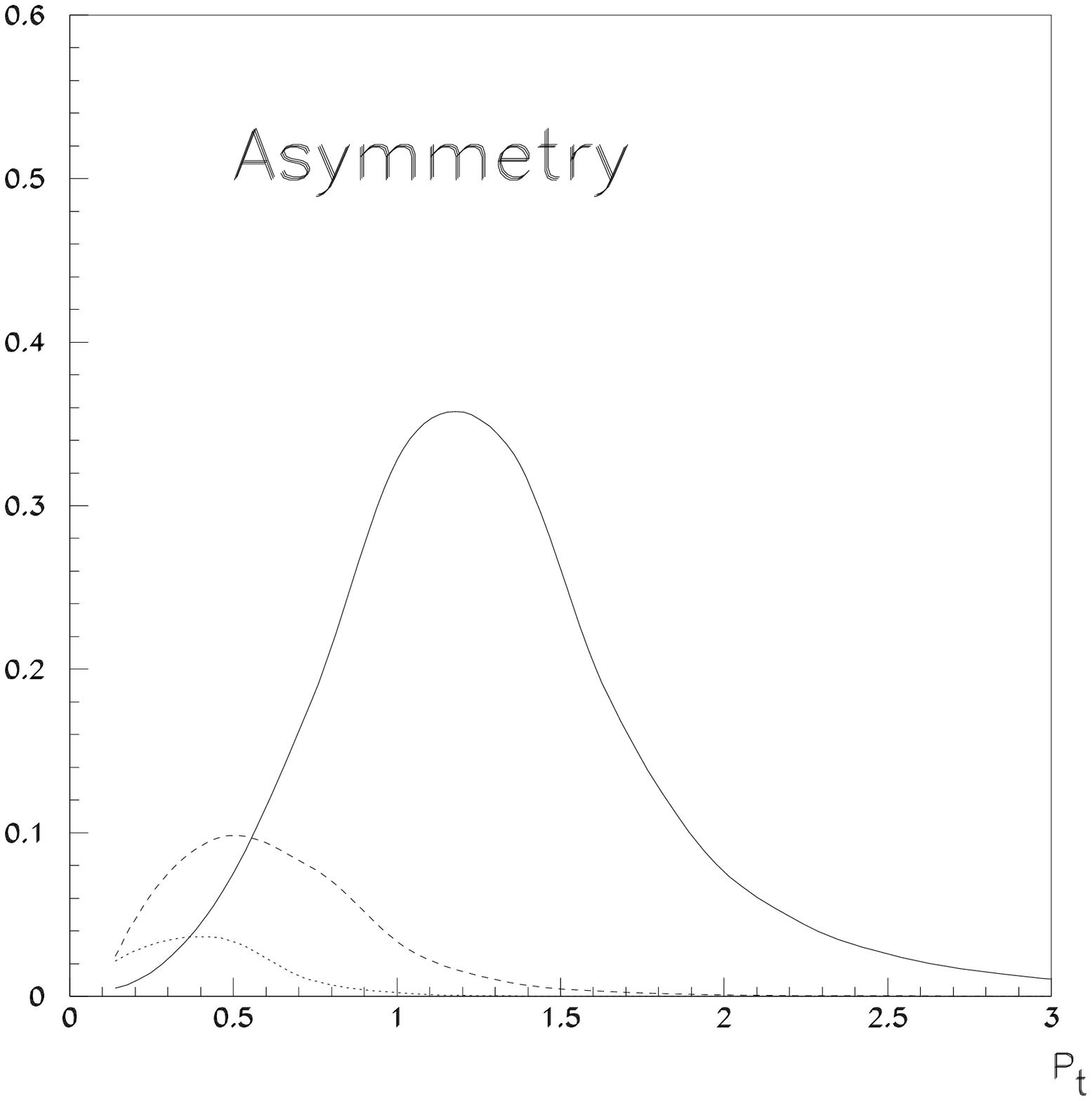,width=6cm}
\caption{\it The instanton contribution to  the  single
spin asymmetry for pion production as a function of $x_F$ and $p_t=|l_\bot|
$.
Solid line is for $x_F=0.9$, dashed line is for $x_F=0.6$ and
dotted line is for $x_F=0.3$. }
\end{minipage}
\end{figure}
The value of the asymmetry is rather large $A\approx 30\%$
in the large $x_F$ and $p_\bot$ region. The magnitude and sign
of the asymmetry  of $\pi^+$ mesons is in qualitative agreement
with the experimental data \ct{E704}. At the same time the negative
and smaller polarization of d-quark in comparison with u-
quark polarization in proton should lead to the negative and small positive
SSA for the $\pi^-$-  and $\pi^0$- meson production, respectively.
This feature also was observed by the E704 Collaboration.
For more detailed comparison
with total set of data one should include in the calculation the
u-and d-quark
distribution functions in the polarized and unpolarized nucleon  and take
the modern
result for the density of instanton from lattice calculations \ct{lattice}.
This will be the subject for a forthcoming paper.

We should stress that  the  instantons give the large SSA at
{\it large} transfer momentum. The scale of the transfer momentum
where one can expect the large SSA in instanton approach is determined
by  the average size of the instanton in QCD vacuum. This size is much
smaller than the confinement size. Therefore the typical
values of the transfer momentum in instanton--induced quark
fragmentation to hadrons are substantially larger than the usual
value $p_\bot\approx 0.2$ GeV related to the confinement scale.
It is this $p_t$ dependence of SSA that poses one of the main
problems in the most attempts to explain the phenomena.
One can also easy understand the origin of the observed enhancement of
the SSA at large $x_F$ region. Indeed the value of the SSA is determined
by product of the  imaginary part of the diagram in Fig.1b and real part
of the diagram in Fig.1a.
The imaginary part of diagram in Fig.1b is proportional to the virtuality
of quark in Fig.1b coming into the instanton vertex \re{form3}. This
virtuality is $k_1^2\approx p_\bot^2/(x_F(1-x_F))$. At the same time
the real part of the diagram in Fig.1a is small at low $x^F$ due to
form factor $F(k_2^2)$, where $k_2^2\approx -p_\bot^2/x_F$.
As result the instanton approach predicts the large SSA only
in large $x_F$ region. This prediction is confirmed by the data \ct{E704}.

It should be mentioned  that instanton (antiinstanton) transition defines the
 particular time direction
\footnote{ The instanton describes the subbarier transition
in time from $-\infty$ to $+\infty$ while the antiinstanton is
the transition in opposite direction.}
 and therefore the possible connection of the instanton
mechanism of the SSA with T-odd fragmentation functions \ct{mulders}
should be clarified.
Recently, a large azimuthal asymmetry in semi-inclusive polarized
electroproduction of pions was observed at HERMES \ct{HERMESas}.
It can be shown that the instanton mechanism for single spin asymmetries
suggested in our paper allows to explain these data \ct{azim} as well.

In summary, the instanton--induced contribution to  quark--quark
scattering amplitude
leads  to a large quark single--spin asymmetry at high energy
and large momentum transfer.
The origin of SSA is in the large imaginary part of the instanton--induced
amplitudes in the time--like region of quark virtuality.
This is related to the  quasiclassical origin of the instanton which
stems from the fact that it is a soliton--like solution  of the QCD
classical equation of motion in  {\it Euclidean} space--time.
We should also emphasize  that appearance of the imaginary part
in the  quark-quark scattering amplitudes which include the
quark lines with time-like momenta
is the common feature  of the instanton--induced processes.
The consideration of the different manifestations of this phenomenon
in polarized and unpolarized lepton-hadron and hadron-hadron interactions
will be the subject of forthcoming papers.

\section*{Acknowledgements}
\vspace{1mm}
\noindent
The  author  is
thankful to M.Anselmino, A. De Roeck, A.E.Dorokhov,
S.B.Gerasimov, E.A.Kuraev, A.V.Efremov,  E.Leader,  M.G.Ryskin and V.Vento
for helpful
discussions. He also very grateful to Prof.A.Di Giacomo for
the warm hospitality at Pisa University where this paper was
started.
The work  was supported in part by the Heisenberg--Landau
and INFN-BLTP JINR exchange programs.

\end{document}